\documentclass[a4paper,12pt]{article}
\usepackage{graphicx}
\usepackage{pdfsync}
\usepackage{color}

\def\l{\left}
\def\r{\right}

\newtheorem{thm}{Theorem}
\newtheorem{cor}{Corollary}
\newtheorem{prop}{Proposition}
\newtheorem{lem}{Lemma}

\let\fd=\rightarrow

\let\eps=\epsilon

\let\dps=\displaystyle

\let\ms=\medskip
\let\mb=\medbreak

\let\bb=\bigbreak

\def\f#1{{\msb F}_{#1}}
\def\pt#1{\left(#1\right)}

\font\tenmsb=msbm10 scaled \magstep1
\font\sevenmsb=msbm9
\font\fivemsb=msbm7
\newfam\msbfam
\def\msb{\fam\msbfam\tenmsb}%
\textfont\msbfam=\tenmsb \scriptfont\msbfam=\sevenmsb%
\scriptscriptfont\msbfam=\fivemsb%

\def\ie{ that is }

\begin{document}

\title{Error-Correction Capability  of Reed-Muller codes}
\author{St\'ephanie Dib, Fran\c cois Rodier \thanks{Aix Marseille Universit\'e, CNRS, Centrale Marseille, Institut de Math\'ematiques de Marseille, UMR 7373, 13288 Marseille, France, \tt stephania.dib@gmail.com, francois.rodier@univ-amu.fr}}
\date{}
\maketitle

\begin{abstract}
We present an asymptotic limit between correctable and uncorrectable errors  on the Reed-Muller codes of any order. This limit is theoretical and does not depend of any decoding algorithm.
\end{abstract}


\section{Introduction}

Let $\f2$ be the field with 2 elements, and
let $RM=RM(n,r)$ be the Reed-Muller  code of length $2^n$ and of order $r$ that is the set of Boolean function with $n$ variables of algebraic degree not more than $r$.

Building a code is important, but we must think about how many words we can decode.
Usually, we content ourselves of the fact that errors of weights less than half of the minimum distance can be corrected in a unique manner.
{So decoding an error correcting code beyond half of the minimum distance has been a challenge for  the one who study error correcting codes.}
In fact experiments show that a maximum likelihood decoding can decode many more words.

Here we propose a theoretical bound for decoding almost all errors of Reed-Muller code on any order  by this method of decoding.
Indeed, the decoder will often be able to recover the correct codeword using an algorithm that generates for each received word the  closest codeword even if the received word is more distant than half of the minimum distance.
On the contrary, when the number
of errors exceeds a certain value, the received vector will be rarely closer to
the correct codeword than to any other one.
Here we give a proof for that.

It is interesting to compare that fact with the phenomenon of concentration of the nonlinearity of Boolean functions which have been studied by several authors (\cite{di1, di2, ls, ro1, ro2, sc}.
The $r$-nonlinearity  of a Boolean function {$f$  denoted $NL_r(f) $} is its Hamming distance to the set of Boolean functions with $n$ variables of algebraic degree not more than $r$.
Claude Carlet \cite{ca}, proved that the density
of the set of Boolean functions satisfying
$$NL_r(f) > 2^{n-1} - c \sqrt{2^{n-1}\pt{n \atop r}\log 2}$$
tends to 1 when $n$ tends to infinity, if $c > 1$.
The authors of the present paper  proved 
a concentration of the nonlinearities of almost all Boolean functions
around 
\begin{equation}
\label{valcri}
 2^{n-1} -  \sqrt{2^{n-1}\pt{n \atop r}\log 2} 
\end{equation}
when $r\le 2$ but missed the greater values by lack of knowledge of weight distributions (\cite{di1, di2,  ro1, ro2}.
Kai-Uwe Schmidt generalized this result for all $r$ thanks to a result of Kaufman, Lovett, and Porat \cite{klp} helping him to find a better bound for the weights of a RM code \cite{sc}.

{On the other hand, Helleseth, Klove and   Levenshtein  in the paper {\sl Error correction
capability of binary linear codes}  \cite{hkl} study  order 1 or 2 Reed-Muller codes and they show that almost all the words are decodable up to the same bound as (\ref{valcri})  and almost all words are not decodable beyond this bound.
For that, they use the monotone structure of correctable and uncorrectable errors.}
St\'ephanie Dib \cite[Chapter 3]{di2}  proved by the same method as for the  concentration of the nonlinearities of almost all Boolean functions that
{the bound for correcting most of the values of codewords for 1-order RM codes was given by 
(\ref{valcri}).}

We show here that the value given in 
 (\ref{valcri}) is also the bound for correcting most of the values of codewords for RM codes for any order.
For RM codes, the present work improves the paper by Helleseth et al.   \cite{hkl} where they just prove the fact that  the codes $RM(n,r)$ are asymptotically optimal for $r=1$ (cf. note after inequality (54) of  \cite{hkl}) or $r=2$ (example 7  of \cite{hkl}). 

\parindent=0mm

\section{Presentation}
 
Let $d(e,f)$ be the Hamming distance between the elements $e$ and $f$ in $\f2^{2^n}$.
We denote by $wt(e)$ the weight of an element $e$ in $\f2^{2^n}$. 
Let $C$ be a linear code of length $m$, of dimension $k$.
The Reed-Muller  code of length $2^n$ and of order $r$ has dimension $\sum_0^r \pt{n \atop r}$ and minimum distance $2^{n-r}$.

\subsection{Correctable and uncorrectable errors}

Let $\f {2}^{2^{n}}$ be the set of all binary vectors of length 2$^n$. For  any vector $f \in \f {2}^{2^{n}}$, the set
\begin{eqnarray*}
f+ C =\{f+g\mid g\in  C \}
\end{eqnarray*}
is called a coset of $ C $ and contains $2^k$ vectors.
One can easily check that two cosets are either disjoint or coincide. This means
\begin{eqnarray*}
f \in h+ C  \Longrightarrow f+ C =h+ C .
\end{eqnarray*}
Therefore, the set $\f {2}^{2^n}$ can be partitioned into $2^{2^n-k}$ cosets of $ C $:
\begin{eqnarray*}
\f {2}^{2^{n}}=\bigcup_{i=0}^{2^{2^n-k}-1}\l(f_i+ C \r),\hspace{2cm} f_i \in \f {2}^{2^{n}}
\end{eqnarray*}
where $\l(f_i+ C \r)\cap\l(f_j+ C \r)=\emptyset$ for $i\neq j$. 

If you send a word $g$ and the decoder receive the word $h$, we will call $e=g-h$ the error.
Thus, the possible error vectors are the vectors in the coset containing $h$. In maximum-likelihood decoding, the decoder's strategy is, given $h$, to choose a minimum weight vector $e$ in $h+ C $, and to decode $h$ as $h-e$. 

{The minimum weight vector in a coset is called the coset leader, and when there is more than one vector of minimum weight in a coset, any one of them can be selected as the coset leader. }

We denote the
set of all coset leaders by $E_ 0(C)$ (note that $\#E_0( C) = 2^{2^n-k}$).
The elements of $E_ 0 (C)$ are called correctable 
errors, and the elements of $E_1(C) = \f2^n- E_0(C)$ are called
uncorrectable errors.
Only coset leaders are correctable errors, which means that $2^{2^n-k}$ errors can be corrected with this decoding.

A codeword is an  unambiguous correctable error if it is a coset leader, and it is the only vector  of minimum weight in this coset.

\begin{prop}.

The following statements are equivalent.
\begin{description}
\item[ 1-] A codeword  $e$ is an  unambiguous  correctable error;
\item[ 2-]  $\forall e'\in e+C$ if $e\ne e'$ then $wt(e)< wt(e')$;
\item[ 3-] $ \forall g\in  C-\{0\},\  wt(e)< wt(g+e) $;
\item[ 4-] $ \forall g\in  C-\{0\},\  d(e,0)< d(g,e)$.
\end{description}

\end{prop}

{\sl Proof}

The first assertion implies the second because
if $e'\in e+C$ and $e\ne e'$  then   $e'$ is not the coset leader, so $wt(e)< wt(e')$.

The second   assertion implies the first because
if $e'\in e+C$ and $e\ne e'$  then $wt(e)< wt(e')$ so $e'$ is not the coset leader and $e$ is the only vector  of minimum weight in this coset.

The other statement are clear.

\subsection{The probability}

We take $\f2^{2^n}$ as the probability space.
We endow it with the uniform probability $P$.

\section{The results}

Let
$\lambda_n= c\times 2^{n/2}\sqrt{2\pt{n\atop r} \log 2}$ and $\delta=2^{n-1} - \lambda_n/2$ where $c$ is a positive real.

\bb

We will show that
if  $c>1$ then almost all error of weight smaller than 
$\delta$
are correctable, when $n$ tends to infinity.
And that
if $c<1$ then almost all error of weight higher than 
$\delta$
are uncorrectable, when $n$ tends to infinity.
More precisely we will show the following two theorems.

\begin{thm}.
\label{thm1}
Let $c>1$. Then
$$P_{wt(e)\le \delta}\Big(d(e,0)<d(e,g) \hbox{ for all $g\in RM(r)-0$}\Big) \fd 1 \hbox{ when }n\fd\infty.$$ 
\end{thm}
and 
\begin{thm}.
\label{thm2}
Let $c<1$. Then
$$\displaylines{
P_{wt(e)\ge \delta}\Big( \hbox{there exists $g\in RM(r)-0$ such that }d(e,0)\ge d(e,g)\Big) \fd 1
\hfill\cr\hfill \hbox{ when }n\fd\infty.
}$$ 

\end{thm}

\bb

\section{Proof of the Theorem \ref{thm1}. Decoding a large number of errors}

We intend to prove that almost all error of weight smaller than $\delta$ for  $c>1$
are   correctable, when $n$ tends to infinity.
It is enough to prove
$$P_{wt(e)\le \delta}\Big(d(e,0)<d(e,g) \hbox{ for all $g\in RM(r)-0$}\Big) \fd 1 \hbox{ when }n\fd\infty.$$

{We have just to show}
$$P_{wt(e)\le \delta}\Big(\delta< d(e,g) \hbox{ for all $g\in RM(r)-0$}\Big) \fd 1 \hbox{ when }n\fd\infty$$

or
$$P_{wt(e)\le \delta}\Big( \exists g\in RM(r)-0,\  \delta\ge d(e,g) \Big) \fd 0 \hbox{ when }n\fd\infty$$
that is
$$P_{wt(e)\le \delta}\pt{ \bigcup_{g\in RM(r)-0}  \Big(\delta\ge d(e,g) \Big) } \fd 0 \hbox{ when }n\fd\infty.$$
It is enough to prove that
$$\sum_{g\in RM(r)-0} P_{wt(e)\le \delta}\pt{  \delta\ge d(e,g)  } \fd 0 \hbox{ when }n\fd\infty.$$

By expressing the conditional probabilities we have to show that
$$ \sum_{g\in RM(r)-0} {P\bigg(\Big({d(e,0)}\le \delta\Big)\cap \Big({ d(e,g)\le \delta }\Big)  \bigg) 
 \over P\Big({d(e,0)}\le \delta\Big)}
  \fd 0 \hbox{ when }n\fd\infty.$$

Let $B_\delta(g)$ be the ball of center $g$ and of radius  $\delta$ that is the set of $e$ such that $d(e,g)\le \delta$.
The event $B_\delta(g)$ is the set of words $f$ in $\f2^{2^n}$ such $f\in B_\delta(g)$, \ie $d(f,g)\le \delta$.

\bb

Hence    Theorem \ref{thm1} is a consequence  of the following proposition.

\begin{prop}.
If  $c>1$ then
$$ \sum_{g\in RM(r)-0} {P\Big( {B_\delta(0)} \cap  {B_\delta(g)}\Big) 
 \over P\pt{B_\delta(0)}}
  \fd 0 \hbox{ when }n\fd\infty$$
\end{prop}

 Before the proof of this Proposition we have to evaluate the terms in the sum.
 
 \begin{lem}.
 \label{inters}
For every real $s$, one has
  \begin{eqnarray*}
P\Big( {B_\delta(0)} \cap  {B_\delta(g)}\Big) 
&\le& \exp\left( 2{s^2} \big(2^{n} - wt(g) \big)-2s\lambda \right).
\end{eqnarray*}

\end{lem}

{\sl Proof}.

Replace $\delta$ by its value.
 \begin{eqnarray*}
P\Big( {B_\delta(0)} \cap  {B_\delta(g)}\Big) 
  &=& P\Big( \pt{wt(f)\le \delta} \cap \pt { wt(f+g)\le \delta}\Big) \\
  &=& P\Big( \big( 2^{n-1}-wt(f)\ge \lambda/2\big) \cap  \big(2^{n-1}- wt(f+g)\ge \lambda/2\big)\Big) 
\end{eqnarray*}
One knows that
$$2^{n }-2wt(f) =\sum_{x\in\f2^n} (-1)^{f(x)}, \quad 2^{n }-2wt(f+g) =\sum_{x\in\f2^n}  (-1)^{f(x)+g(x)}.$$
Hence this gives using Markov's inequality:
\begin{eqnarray*}
\lefteqn{P\Big( {B_\delta(0)} \cap  {B_\delta(g)}\Big) } \hspace{10mm} \\
  &=& P\left( \pt {\sum_{x\in\f2^n}  (-1)^{f(x)}\ge \lambda} \cap  \pt{\sum_{x\in\f2^n}  (-1)^{f(x)+g(x)}\ge \lambda}\right)\\
  &=& P\left( \pt {\exp\pt{s\sum_{x\in\f2^n}  (-1)^{f(x)}}\ge\exp (s \lambda)} \right.\cap  \\
  &&\hspace{35mm}\left.\pt{\exp\pt{s\sum_{x\in\f2^n}  (-1)^{f(x)+g(x)}}\ge \exp(s\lambda)}\right)\\
  &\le& E\left( \exp\Big(s\sum_{x\in\f2^n}  (-1)^{f(x)}\Big)   \exp \Big(s\sum_{x\in\f2^n}  (-1)^{f(x)+g(x)}\Big) \right)\Bigg/ \exp(s\lambda)^2\\
\end{eqnarray*}
Since the {random values} $f(x)$ are independant 
    \begin{eqnarray*}
P\Big( {B_\delta(0)} \cap  {B_\delta(g)}\Big) 
  &\le& E\left(\exp\Big(\sum_{x\in\f2^n} s  (-1)^{f(x)}\Big(1+ (-1)^{ g(x)}\Big) \right)\Bigg/ \exp(s\lambda)^2\\
  &\le& \prod _{x\in\f2^n}E\left(\exp\Big(s  (-1)^{f(x)}\Big(1+ (-1)^{ g(x)}\Big) \right)\Bigg/ \exp(s\lambda)^2\\
\end{eqnarray*}
Because the {random values} $f(x)$ takes the values $\pm1$ with probability $1/2$, the calculation of the expectation gives
   \begin{eqnarray*}
P\Big( {B_\delta(0)} \cap  {B_\delta(g)}\Big) 
 &\le& \prod _{x\in\f2^n}\cosh\left( s \Big(1+ (-1)^{ g(x)} \Big) \right)\Bigg/ \exp(s\lambda)^2
\end{eqnarray*}
As
$\cosh(t)\le \exp(t^2/2)$
  \begin{eqnarray*}
P\Big( {B_\delta(0)} \cap  {B_\delta(g)}\Big) 
  &\le& \prod _{x\in\f2^n}\exp\left( s^2 \Big(1+ (-1)^{ g(x)} \Big)^2/2 \right)\Bigg/ \exp(s\lambda)^2\\
  &\le& \exp\left( {s^2} \Big(2^n+ \sum _{\f2^n}(-1)^{ g(x)} \Big) \right)\Bigg/ \exp(s\lambda)^2\\
  &\le& \exp\left( 2{s^2} \big(2^{n} - wt(g) \big)-2s\lambda \right).
\end{eqnarray*}

  \subsection{Case where the distances are close to $2^{n-1}$.}
We give a bound for $P\Big( {B_\delta(0)} \cap  {B_\delta(g)}\Big) $ when the distance to 0 of the center $ g$ is rather close to $2^{n-1}$.
\begin{lem}.
If
  $$|2^{n-1}-d(g,0)| \le 2^{n-1}/\pt{n\atop r}$$
  then:
  \begin{eqnarray*}
P\Big( {B_\delta(0)} \cap  {B_\delta(g)}\Big) 
&\le&\dps {1\over 2^{c ^2 {2}\pt{ \pt{n\atop r}-1}}} .
\end{eqnarray*}

  \end{lem}  
  
  {\sl Proof}.
  
  From lemma \ref{inters} we have
 \begin{eqnarray*}
P\Big( {B_\delta(0)} \cap  {B_\delta(g)}\Big) 
  &\le& \exp\bigg( {s^2} \Big(2^n+ 2^{n}-2wt(g) \Big) \bigg)\Bigg/ \exp(s\lambda)^2\\
  &\le& \exp\bigg( {s^2}2^n \Big(1+ 1\Big/\textstyle\pt{n\atop r} \Big) \bigg)\Bigg/ \exp(s\lambda)^2\\
\end{eqnarray*}

We take $s=\lambda/2^n$.
\begin{eqnarray*}
P\Big( {B_\delta(0)} \cap  {B_\delta(g)}\Big) 
  &\le& \exp\bigg( \lambda^2 2^{-n} \pt{1+ 1\Big/{\textstyle\pt{n\atop r} }} \bigg)\bigg/ \exp(\lambda^2/2^{n-1})\\
  &\le& \exp\bigg(2 c^2  {\textstyle\pt{n\atop r} \log 2} \pt{1+ 1\Big/ {\textstyle\pt{n\atop r}}  }\bigg)\bigg/ \exp\Big( 4c^2  {\textstyle\pt{n\atop r} \log 2}\Big)
  \end{eqnarray*}

Simplifying the two members of this fraction by $\exp\left(2 c^2  {\textstyle\pt{n\atop r} \log 2}\right) $ you get
\begin{eqnarray*}
P\Big( {B_\delta(0)} \cap  {B_\delta(g)}\Big) 
  &\le&{   \exp\left(2 c^2 {  \log 2}  \right)
   \over \exp(2 c^2  {\pt{n\atop r} \log 2})}
  \le{   2^{2  c^2 }   \over 2^ {c^2\times {2\pt{n\atop r}}}}
\end{eqnarray*}

\subsection{Case where the distances are away from $2^{n-1}$.}

We use the follwing lemma, which is an application of a result by Kaufman, Lovett, and Porat \cite{klp}.
 
\begin{lem}.
Let  $\alpha$ be a strictly positive real number.
The number $B_{r,n}$ of functions $g$ in $RM(r, n)$ satisfying
$$|wt(g)-2^{n-1}| \ge 2^{n-1}/\pt{n\atop r}$$
 fulfills
$$B_{r,n}\le 2^{\alpha \pt{n\atop r}}$$
if $n$ is large enough.

\end{lem}

{\sl Proof}

This is shown in the proof of Lemma 3 in K.-U. Schmidt's article \cite[relation (6)]{sc}.
  \bb
We use this lemma to evaluate $\Pi=\sum P\Big( {B_\delta(0)} \cap  {B_\delta(g)}\Big) 
$
where the sum is  on the nonzero  $g$ in $RM(n,r)$ fulfilling
$$|wt(g)-2^{n-1}| \ge 2^{n-1}\Big/\pt{n\atop r}.$$
\begin{lem}.
Let $\alpha$ be a strictly positive real number. Then
$$\Pi<  2^{\alpha \pt{n\atop r}} 2^{-{c^2\over 1-2^{-r}}\pt{n\atop r}}$$

\end{lem}
{\sl Proof}.

From lemma 1, for all $s$, we have
  \begin{eqnarray*}
P\Big( {B_\delta(0)} \cap  {B_\delta(g)}\Big) 
&\le& \exp\left( 2{s^2} \big(2^{n} -  wt(g) \big)-2s\lambda \right)
  \end{eqnarray*}
  
  Let us take $\dps s={\lambda\over 2^{n+1} - 2wt(g) }$. We have, expressing the value of $\lambda$ and noting that $wt(g)$ is not less than the minimum distance $2^{n-r}$ of $RM(n,r)$:
\begin{eqnarray*}
P\Big( {B_\delta(0)} \cap  {B_\delta(g)}\Big) 
&\le& \exp\left( - {\lambda^2 \over 2^{n+1 } -  2wt(g)} \right)\\
&\le& \exp\left(  - {c^2\times 2^{n+1} {\pt{n\atop r} \log 2} \over 2^{n+1 } -   2^{n-r+1}} \right)\\
&\le& 2^{  - {c^2\times   { \pt{n\atop r}} \over 1 -   2^{-r}} }.
\end{eqnarray*}

Therefore
$$\Pi \le B_{r,n}   2^{-{c^2\over 1-2^{-r}}\pt{n\atop r}} \le  2^{\alpha \pt{n\atop r}} 2^{-{c^2\over 1-2^{-r}}\pt{n\atop r}}.$$

\subsection{Evaluation of $P\pt{B_\delta(0)}$}

\begin{prop}.\label{Pboule}
Let $r$ be a fixed integer, $\delta=2^{n-1}-c\,\sqrt{2^{n-1}\pt{{n}\atop{r}}\log2}$ where
$c$ is a positive constant. We have
\begin{equation}
 P\pt{B_\delta(0)}
 \label{Bdelta}
={1\over 2\pi}{ 2^{-  c^2  {\pt{n\atop r}}}\over  2c  \sqrt{\pt{n\atop r} \log 2}} (1+o(1))
\end{equation}
when $n$ tends to infinity.

\end{prop}

This is proved in St\'ephanie Dib's thesis \cite{di2}.
We recall briefly the proof for completeness.

The following lemma (see \cite[lemma 1]{ca}) gives well-known asymptotic estimate of the sum of binomial coefficients.
\begin{lem}.
Let $n$ be a positive integer and $k\leq n$. Then
$$
\sum_{i=0}^{k}\pt{2n\atop i}<2^{2n}\cdot\exp\left(-\frac{(n-k)^2}{n}\right).\label{sombi}$$
\end{lem}

When $k$ is sufficiently close to $n$, the following lemma (see \cite[chapter IX, (9.98)]{gkp}, \cite[chapter VII]{fe}) gives an asymptotic estimation for $\pt{2n\atop k}$:
\begin{lem}.
Let $n$ be a positive integer and $|n-k|\leq n^{\frac{5}{8}}$. Then 
\begin{equation}
\pt{2n\atop k}= \frac{2^{2n}}{\sqrt{\pi\cdot n}}\cdot\exp\left(-\frac{(n-k)^2}{n}\right)\cdot\left(1+o(1)\right),\label{O&S}
\end{equation}
where the term $o(1)$ is independent of the choice of $k$.
\end{lem}

{\sl Proof of the Proposition.}

The number of Boolean functions whose Hamming distance
to $0$ is bounded from above by some number $\delta$ equals
$\dps\sum_{0\leq i\leq \delta} \pt{2^{n}\atop i}.$
Thus we have
$$
\sum_{0\leq i\leq \delta} \pt{2^{n}\atop i} = \sum_{0\le i < 2^{n-1}-2^{(n-1)\frac{5}{8}} } \pt{2^{n} \atop i}+ \sum_{2^{n-1}-2^{(n-1)\frac{5}{8}}\leq i\leq \delta} \pt{2^{n} \atop i}.
$$
The first sum on the right hand side is taken care of by lemma \ref{sombi} which will show  that it is negligible with respect of the second sum.

To estimate a lower bound of the second sum (which we denote $S$), we use (\ref{O&S})
\begin{eqnarray*}
S
&=&{2^{2^n}\over\sqrt{\pi\,2^{n-1}}}\cdot\left(1+o(1)\right)\cdot
\sum_{2^{n-1}-2^{(n-1)\frac{5}{8}}\leq i\leq \delta}\exp\left(-\frac{(2^{n-1}-i)^2}{2^{n-1}}\right).  \\
\end{eqnarray*}

We use that the function in the sum is monotonous to replace the sum by an integral.
\begin{eqnarray*}
S
&=&{2^{2^n}\over\sqrt{\pi\,2^{n-1}}}\cdot\left(1+o(1)\right)\cdot
\int_{2^{n-1}-2^{(n-1)\frac{5}{8}}+1\leq i\leq \delta }\exp\left(-\frac{(2^{n-1}-i)^2}{2^{n-1}}\right)di.\\
&=&{2^{2^n}\over\sqrt{\pi}}\cdot\left(1+o(1)\right)\cdot 
\int_{{c\,\sqrt{\pt{{n}\atop{r}}\log2} } \leq  v\leq 2^{n-1\over 8}-2^{1-n\over2}}\exp\left(-v^2\right)dv.\\
\end{eqnarray*}
By  \cite[chapter VII, Lemma 2]{fe} and the fact that 
$$c^2\pt{{n}\atop{r}}\log2 - \pt{2^{n-1\over 8}-2^{1-n\over2}}^2\fd-\infty$$
which implies that
$$\int_{ 2^{n-1\over 8}-2^{1-n\over2}}^\infty  \exp\left(-v^2\right)dv
=o\pt{
\int_{{c\,\sqrt{\pt{{n}\atop{r}}\log2} }}^\infty \exp\left(-v^2\right)dv}$$
the last integral is equivalent to
$${\exp\pt{-c^2\ {\pt{{n}\atop{r}}\log2} }\over 2c\,\sqrt{\pt{{n}\atop{r}}\log2} }
={2^{-c^2\ {\pt{{n}\atop{r}}} }\over 2c\,\sqrt{\pt{{n}\atop{r}}\log2} }.
$$
Thus
\begin{eqnarray*}
\sum_{0\leq i\leq \delta} \pt{2^{n}\atop i}
&=&
{ 2^{2^n}\over\sqrt{\pi}}
{2^{-c^2\ {\pt{{n}\atop{r}}} }\over 2c\,\sqrt{\pt{{n}\atop{r}}\log2} }(1+o(1)).
\end{eqnarray*}

\subsection{Proof of Theorem \ref{thm1}}

 Therefore
$$\displaylines{
 \sum_{g\in RM(r)-0} {P\Big( {B_\delta(0)} \cap  {B_\delta(g)}\Big) 
 \over P\pt{B_\delta(0)}} \hfill\cr\hfill
\le O(n^{r/2}) \pt{2^{\pt{n\atop r}}{ 2^{-c ^2 {2}\pt{ \pt{n\atop r}-1}}}2^{  c^2  {\pt{n\atop r}}}
+2^{\alpha \pt{n\atop r}}  2^{-{c^2\over 1-2^{-r}}\pt{n\atop r}} 2^{  c^2  {\pt{n\atop r}}}}.
}
$$

 This tends to 0 because the exponent of 2 is, for the left term
\begin{eqnarray*}
&&\pt{n\atop r} -c ^2 {2}\pt{ \pt{n\atop r}-1} +c^2  {\pt{n\atop r}}
=-\pt{n\atop r}{c ^2  }+2c ^2
 \fd -\infty
\end{eqnarray*}
and for the right term
\begin{eqnarray*}
&&\alpha \pt{n\atop r}-{c^2\over 1-2^{-r}}\pt{n\atop r}+  c^2  {\pt{n\atop r}}
= \pt{n\atop r}\pt{\alpha-{2^{-r}c^2\over 1-2^{-r}} }.
\end{eqnarray*}
So just take
$$\alpha<{2^{-r}c^2\over 1-2^{-r}} $$
so that this term tends to  $-\infty$.

\section{The error correction capability function}
\label{order}

Let
$\eps_C(t)$ the ratio of the number of {correctables} errors  of weight $t$ to the number of {words} of weight $t$.
 {Let us suppose from now on that   the lexicographically
smallest minimum-weight vectors are chosen as the coset
 leaders. This involves only the cosets with several minimum weight vectors that is
  the ambiguous correctable errors.
 Then an important property of this ratio is that
 for any $t$ in the range from half the
 minimum distance to the covering radius, $\eps_C( t)$ decreases with
 the growing $t$ as the next lemma says.
}

\begin{lem}.

For any $[n, k]$ code $C$ and any $t = 0, 1,\dots,n-1$
$$\eps_C(t+1)\le \eps_C(t)$$
with strict inequality for $t_C\le t\le r_C$ where we set $t_C = \lfloor( d_C -1)/2\rfloor$ and denote the
covering radius of C by $r_C$.
\end{lem}

{\sl Proof}

See Helleseth et al. \cite[Lemma 2]{hkl}.
This property is due to the fact that the sets of correctable and uncorrectable errors form
a monotone structure,  (see, for example, \cite[p. 58, Theorem
3.11]{pw}) and a result of Bollobas about shadows \cite[Theorem 3]{bo}

\subsection{A corollary of Theorem \ref{thm1}}

For Reed-Muller codes of  order $r$, that is to say $RM(n,r)$ we take
$$t
=2^{n-1}-c \sqrt{2^{n-1} \pt{n \atop r} \ln2}.$$

\begin{cor}.
 $\hbox{If }c>1, \hbox{ then }\eps_C(t_c) \fd 1 \hbox{ when }n\fd\infty.$ 
\end{cor}

{\sl Proof}.

We know that the ratio of the unambiguous correctable errors (hence also the correctable errors)  of weight smaller than $t{_c}$ to the words  of weight smaller than $t{_c}$ tends to 1 when $n$ tends to infinity. We have to show that the ratio of the correctable errors of weight exactly  $t{_c}$ to the words  of weight exactly  $t{_c}$ tends to 1 when $n$ tends to infinity. 

For an $RM(n,r)$ code, let 
$${ 2^{2^n}\over\sqrt{\pi}} {2^{-c^2\ {\pt{{n}\atop{r}}} }\over 2c\,\sqrt{\pt{{n}\atop{r}}\log2} } = A(c).$$
Let us fix $c_1$ and suppose that $\eps_C(t_{c_1})\not\fd 1$. Then there exists $\eta<1$ such that $\eps_C(t_{c_1})<\eta$ for an infinity of $n$.
If $c_1> c_2 >1 $, then 
among the words of weights  between $t_{c_1} $ and $ t_{c_2}$ there is only at most a proportion $ \eta $ of correctable words as the function $\eps_C$ decreases. From Proposition \ref{Pboule} there are about
\begin{eqnarray*}
\sum_{0\leq i\leq \delta_1} \pt{2^{n}\atop i}
&=&
A(c_1)(1+o(1)).
\end{eqnarray*}
words of weight in $[0,\ t_{c_1}]$
and 
\begin{eqnarray*}
\sum_{\delta_1\leq i\leq \delta_2} \pt{2^{n}\atop i}
&=&
A(c_2)(1+o(1)).
\end{eqnarray*}
words of weight between $t_{c_1}$ et $t_{c_2}$. 
As $A(c_1)=o(A(c_2))$ there are at most
$$A(c_1)(1+o(1))
+\eta A(c_2)(1+o(1))
=\eta A(c_2)(1+o(1))
$$
correctable words of weights $[0,\ t_{c_1}]$, which shows that it is impossible that  almost all words are correctable as says Theorem \ref{thm1}.

\subsection{Proof of the Theorem \ref{thm2}. An asymptotic decoding upper bound}

\bb
{In the case of RM codes we have a simplification of the proof of the Theorem 3 b in \cite{hkl}.}

\begin{prop}.
 $\hbox{If }c<1, \hbox{ then }\eps_C(t_c) \fd 0 \hbox{ when }n\fd\infty.$ 
\end{prop}

For every $t$, one has (cf. Lemma 3 of \cite{hkl})
\begin{eqnarray*}
\eps_C(t) \times  \dps\sum_{i=0}^t \pt{2^n \atop i} 
&\le & \sum_{i=0}^t \eps_C(i)\pt{2^n \atop i} \cr
&=& \hbox{ number of correctable errors  of weight smaller than } t  \cr
&\le &\hbox{  total number of correctable errors   }   \cr
&\le &2^{2^n-k}.
\end{eqnarray*}

\bb

We have, from Proposition \ref{Pboule}
\begin{eqnarray*}
\sum_{i=0}^{t_c} \pt{2^n \atop i} 
&=& \# B_{t_c} 
= {2^{2^n -c^2\pt{n \atop r}} \over 2c\sqrt{\pi \pt{n \atop r} \ln2}}(1+o(1)).
\end{eqnarray*}
Whence
\begin{eqnarray*}
\eps_C(t_c)  
&\le& {\dps 2^{2^n-k}\over \dps  \sum_{i=0}^{t_c} \pt{2^n \atop i} }\cr
\noalign{\ms}
&= &{2c\sqrt{\pi \pt{n \atop r} \ln2}\over 2^{ 2^n -c^2\pt{n \atop r}}\times \dps 2^{\sum_{i=0}^r \pt{n \atop i}- 2^n} }  (1+o(1))\cr
\noalign{\ms}
&=& {2c\sqrt{\pi \pt{n \atop r} \ln2}\over 2^{\sum_{i=0}^r \pt{n \atop i} -c^2\pt{n \atop r}} }(1+o(1)).
\end{eqnarray*}
If $c < 1$,  when $n \fd \infty$ then 
the denominator tends toward infinity, so $\eps_C(t_c) \fd 0$.

\mb{\bf Remark. }

This {proposition} is still true if we take $t_c=2^{n-1}-c \sqrt{2^{n-1} k \ln2}$ with $c<1$ if $k\le 2^{(n-1)/4}$ (to be able to use Proposition \ref{Pboule}).

\subsection{End of the proof of Theorem \ref{thm2}.}

Remark that  the statement of Theorem \ref{thm2} does not involves  ambiguously correctable errors.
So we have the choice of ambiguous correctable errors, and we can choose  
 the lexicographically
smallest minimum-weight vectors as the coset
 leaders  as in the beginning of the section \ref{order}.

Then one has from the last proposition
$$P_{wt(e)\ge \delta}\Big( e \hbox{ is correctable}\Big)\le  \eps(\delta) \fd 0$$
when $n$ tends to infinity.
Therefore 
$$P_{wt(e)\ge \delta}\Big( e \hbox{ is unambiguously correctable}\Big) \fd 0$$
which means that Theorem \ref{thm2} is true.

\subsection{Asymptotically optimality of  RM codes}

A sequence $(C_m)_m$ of $[m, k]$ codes where $k = o(m)$  as $m\fd\infty$
is
called asymptotically optimal if for any fixed $\eps$, $0 < \eps < 1$
 $$m - 2t_{C_m} (\eps) \sim \sqrt{mk \ln 4}$$
where the
 error correction capability
function $t_C (\eps)$ is the maximum $t$ such that
 $\eps_C(t)\ge\eps$.
 
 \begin{thm}.
The sequence $RM(n,r)_n$ is asymptotically optimal.
\end{thm}

{\sl Proof}.

Let us take $\eps$ and try to find $ t_C(\eps)$. 

Let $c<1$.

If $2{t_c}=2^n-c\sqrt{2^n k \log 4}\ $ then the words of weight ${t_c}$ are almost all uncorrectable, therefore ${\eps_{C_{2^n}}({t_c})\fd 0}$ as $n\fd\infty$. And we have $\eps_{C_{2^n}}({t_c})<\eps$ for $n$ big enough  (and consequently ${t_c}\ge t_{C_{2^n}}(\eps)$).

As a result $t_{C_{2^n}}(\eps)\le 2^{n-1}- c\sqrt{2^{n-1} k \log 4}$.

\mb
Let now $c>1$.

If $2{t_c}=2^n-c\sqrt{2^n k \log 4}\ $  then the words of weight $ {t_c} $ are almost all correctable, therefore ${\eps_{C_{2^n}}({t_c})\fd 1}$ as $n\fd\infty$. And we have $\eps_{C_{2^n}}({t_c})>\eps$  for $n$ big enough  (and consequently ${t_c}<t_{C_{2^n}}(\eps)$).

As a result, if $c_1<1< c_2$, one has  
$$2^n- c_2\sqrt{2^n k \log 4}<2 t_{C_{2^n}}(\eps)< 2^{n}- c_1\sqrt{2^{n} k \log 4}$$
or
$$ c_1\sqrt{2^n k \log 4}<2^n-2 t_{C_{2^n}}(\eps)<  c_2\sqrt{2^{n} k \log 4}$$
or
$$ c_1 <{2^n-2 t_{C_{2^n}}(\eps)\over \sqrt{2^n k \log 4}}<  c_2.$$
As $c_1$ and $c_2$ may be as close to 1 as we wish, we have
$${2^n-2 t_{C_{2^n}}(\eps)\over \sqrt{2^n k \log 4}}\fd1\quad \hbox{{when}}\quad n\fd\infty.$$

\end{document}